# THE COURSE DESIGN TO DEVELOP META-COGNITIVE SKILLS FOR COLLABORATIVE LEARNING THROUGH TOOL-ASSISTED DISCOURSE ANALYSIS

Yoshiaki Matsuzawa, Sayaka Tohyama, and Sanshiro Sakai

Shizuoka University
3-5-1 Johoku
Hamamatsu, Shizouka, Japan
e-mail: matsuzawa@inf.shizuoka.ac.jp

## ABSTRACT

This paper presents the course design titled "Learning Management" of which the goal is to "learn collaborative learning" for a first-year undergraduate student. The objective of the class design is to help transform the student's belief of learning from a passive, individual model to an active, collaborative model which is supported by the concept of "Knowledge Building" or "Constructive Interaction". We conducted an empirical study where the students analyzed their own discourse with KBDeX which is the software they used to assist their analysis in the experimental group whereas the students in the control group reflected their project activities in their own way. We examined the transformation of their beliefs through the qualitative analysis of their reports after the course. The results showed that the design led to the transforming of their learning beliefs from "just experiences of the participation of the collaborative learning" to "active contribution for the collaborative knowledge creation". It showed we succeeded in changing the students' preferences about collaborative learning from negative to positive as well.

## INTRODUCTION

In the past decade, improvement of ICT has changed the nature of how work is conducted and the meaning of social relationships (Binkley et al., 2012). Ways of working are trending toward creating new knowledge, requiring a transformation in ways of learning. Industries have declared a requirement for professionals with the skills for collective knowledge creation, which led to the 'Assessment & Teaching of 21st-Century Skills' project (Binkley et al., 2012).

'Knowledge building' (Bereiter, 2002) is a leading model for developing 21st century skills (Scardamalia et al., 2012), and one of the three prominent models given in Paavola's knowledge-creation metaphor (Paavola et al., 2004). In the knowledge-building community, the concept of "embedded and transformative assessment" has been proposed for the assessment of knowledge-building activities (Scardamalia, 2002). The concept is described as "The community engaging in its own internal assessment, which is both more fine-tuned and rigorous than external assessment, and serves to ensure that the community's work will exceed the expectations of external assessors" (Scardamalia, 2002).

The final goal of this research is to present a tool to capture knowledge-building phenomena as an emergent and embedded process. Hence, we have developed a methodology and supporting tools for discourse analysis in collaborative learning for both researchers and learners (Oshima et al., 2012; Matsuzawa et al., 2012). These studies show that the tool can reveal similar conclusions to in-depth qualitative analysis. In this paper, we expanded the application of the tools to the mainstream of the course design titled "Learning Management" of which the goal is to "learn collaborative learning" for a first-year undergraduate student.

## RELATED WORK

It is well known that the skill of meta learning is a fundamental ability in learning (Palincsar & Brown,1984; Scardamalia et al, 1984). The theory of 'constructive interaction' (Miyake, 1986) also suggests the importance of the meta point of view. She pointed out that "monitors" often give key information to solve problems by taking the meta point of view. Hence, it is difficult to describe how to design curricula of learning meta learning skills. Bransford et al. (1999) suggested that the learning of meta learning skills could not be separated from learning of content when designing learning environments. It seems designing curriculum which requires students to use meta learning skills in learning of content is essential for students. Collaborative learning enables students to reach deeper understanding than students learning alone (H-Silver et al., 2013). Meta learning is also important for students in such collaborative learning environments. Barron (2003) suggested that the students within high performance groups got higher scores after a group activity even though there was no difference between the students' achievements before the group activity. Damsa et al. (2010) claimed that the skill of "shared epistemic agency" predicts the quality of group products through comparing a group which produced sophisticated product to a group which made normal product. However, how to grow students' meta learning skills in collaborative learning had not been established.

A related study concerning visualization of the knowledge-creation metaphor for CSCL (Computer-supported Collaborative Learning) discourse analysis was conducted by Suthers et al. (2010), who proposed a unique framework for conceptualizing and representing distributed interaction. Shaffer et al. (2009) proposed an original framework to assess 21st century skills using what is call epistemic network analysis (ENA). While the purpose and approach is similar to that of our research. Lee et al. (2006) and Aalst and Chan (2007) investigated self-assessments in knowledge-building environments using e-portfolio online notes and visualized data from the analytic toolkit (ATK). ATK provides basic quantitative metrics for online knowledge-building activities, such as the number of online notes created and read, and the percentage of linked notes and keywords.

Several studies have been conducted to capture the nature of knowledge building using the SNA (Social Network Analysis) approach (Laat et al., 2007; Zhang et al., 2009; Yoon, 2011). However, in the three successful SNA applications, researchers created networks based on learner interaction. A similar approach as our approach has been conducted for general communication on the web by Gloor's group, who proposed a SNA-supporting tool that shows a temporal visualization of the network evolution over time (Gloor and Zhao, 2004). They also applied LSA (Latent Semantic Analysis), and semantic SNA (Gloor et al., 2009), and clustering technologies to the network sciences (Fuehres et al., 2011). However, these are limited to a shallow level of text mining, as a result there is no successful model to reveal the semantic nature of the educational environment.

## METHOD

### Classroom Descriptions and Experimental Study Design

We conducted an empirical study in a first-year undergraduate course titled "Learning Management". The objective of the course was to "learn learning" through engaging in knowledge building activities.

In the entire 15 weeks of the course, students were engaged in two projects each of which took 5 weeks. The students were given a knowledge building inquiry such as "How was privacy saved from security cameras", or "Design a new workshop to teach something". A two week reflection session was scheduled after each project was finished.

Experimental research design was applied for the reflection session. In the 2012 class, students were required to conduct their own discourse analysis, completing the specific analysis format and KBDeX. The details are described in the following session. In the 2010 class, students were required to reflect their project activities in their own way. In other words, the 2012 class was an experimental group with tools, whereas the 2010 class was a control group without tools.

### KBDeX: Tool to Assist the Discourse Analysis

In order to support their discourse analysis, KBDeX is provided for the experimental group. KBDeX (Knowledge Building Discourse Explorer) is the tool which supports discourse analysis in collaborative learning from the perspective of complex network science (Oshima et al., 2012). The tool developed by us has intuitive user interface as the previous paper demonstrated (Matsuzawa et al., 2012 ).

KBDeX visualizes network structures of discourse based on a bipartite graph of words × discourse units (e.g., conversation turns, BBS postings, or sentences) (Figure 1). Using discourse data (in .csv format) and a list of target words for bipartite graph creation (a text file) as its input, KBDeX can create visualizations of three different network structures: (1) learners' network structure (top right window in Figure 1), (2) unit network structure (bottom left), and (3) the network structure of the target words (bottom right).

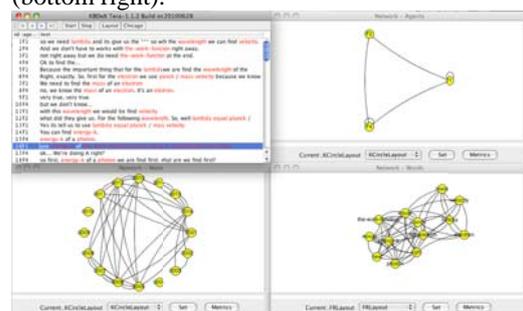

*Figure 1: KBDeX Software Interface*

**The Analysis Sheet**

In the 2012 course, we provided an analysis sheet for students in order to scaffold analyzing their own discourse analysis. The sheet was designed based on our previous study (Matsuzawa, et, al., 2012), which described how discourse analysis was done by students. The sheet includes inquiries for students listed as follows:
1) select and list twenty keywords of the discourse.
2) summarize and list three topics built in the discourse.
3) articulate the discourse into some phase in order to explain the process of the discussion. Tag each phase as three categories (knowledge sharing, knowledge construction, knowledge creation (Aalst, 2009))
4) select and list five important discourse units (notes), and explain a reason for each choice .
5) explain roles and contributions in the discourse for each individual.
6) describe the things you should improve for the next time in collaborative learning.

**Data Collection and Analysis Method**

We analyzed each student's report and questionnaire as described below.

*Report Analysis*

We asked the question: "What is the most impressive thing to change your learning belief" to students in the term report after all classes in the course were finished.

All descriptions for the question were analyzed qualitatively. The eleven coding categories were created by the pedagogical aspects as follows: Knowledge Sharing (SHR), Communication Skills (COM), Idea Diversity (DIV), Controversy (CNT), Argument Elaboration (ELB), Deep Understanding (DEP), Reasoning and Evidences (EVD), Knowledge Creation (CRA), Passive to Active (PTA), Meta Learning (MET), and Collaboration Management (MNG). These descriptions were summarized in Table A1, and Table A2 as an appendix.

Additionally, the answers in the term report about "what kind of activity is ideal in group work" were analyzed qualitatively in another way. RubKB (the five-graded rubric) was made by aspects of students' belief about collaborative knowledge building activity based on ITL Research's rubric of collaboration (ITL Research, 2012; shown in Table A3 as an appendix). We divided the meaning of "shared responsibility" in ITL rubric into "role-sharing (no interaction after sharing roles)" and "exchanging one's own thinking (no improvement hearing other's thinking)" to capture students' beliefs in detail. We divided the original ITL's level 3 into levels 3 and 4 in our RubKB. Also we incorporated the ITL level 1 definition into the RubKB level 1 definition. Please note that the ITL and RubKB level 1 wording is different though the meaning remains the same. The description of our rubric is shown in Table A4 as an appendix. The analysis process was done for the data both in 2010 and 2012 to compare the differences.

*Questionnaire*

We asked students their preferences of learning by questionnaire in order to investigate whether students' attitudes were changed by experiences in the course. Students answered the two questions below in the 5 levels Likert Scale (1-5).
1) Do you like (general) learning?
2) Do you like collaborative learning?
The questionnaire was applied twice, both before taking the course and after finishing the 15 week course to compare how their score changed. We could not compare between years, because this analysis was conducted only for the 2012 class.

**RESULTS**

**Improvement of the students' learning beliefs**

The results of qualitative analysis of reports for the question of their improvement of the learning belief are shown in Table 1. The table has two rows for comparison between the 2010 class without tool analysis and the 2012 class with tools.
Students in the 2010 class described basic collaborative learning features such as Communication Skills, Idea Diversity, and Controversy, whereas students in the 2012 class described many knowledge building terms such as Knowledge Creation and Passive to Active. In 2010, students also referred to characteristics such as Argument Elaboration and Reasoning and Evidence whereas students in 2012 referred to Meta Learning Activities.

The results of the question in the term report "what kind of activity is ideal in group work" are shown in Figure 2. The average score was 2.31 for the 2010 class and 3.20 for the 2012 class. An unpaired t-test has been conducted for comparison. The average score in the 2012 class is significantly higher than that for the 2010 class ($t(81)=4.58$, $p<.01$). The students' term reports from 2012 revealed a higher level of understanding concerning collaborative knowledge building than the students from 2010.

**Questionnaire about preference about learning**

We compared the scores of the questionnaire about the students' preference of learning in 2012. The results are shown in Figure 3. The mean of preference of the general learning was 3.75 before taking the class and 3.66 after taking the class. The mean of preference of the collaborative learning was 2.82 before taking the class and 3.50 after taking the class.

| class | n | SHR | COM | DIV | CNT | ELB | DEP | EVD | CRA | PTA | MET | MNG |
|-------|---|-----|-----|-----|-----|-----|-----|-----|-----|-----|-----|-----|
| 2010 | 49 | 5 | 7 | 10 | 9 | 3 | 3 | 3 | 0 | 1 | 0 | 5 |
| 2012 | 46 | 5 | 3 | 2 | 0 | 0 | 4 | 0 | 8 | 5 | 7 | 2 |

*Table 1: The results of qualitative analysis of reports for question of their improvement of learning belief in comparison between class 2010 without discourse analysis and class 2012 with discourse analysis.*

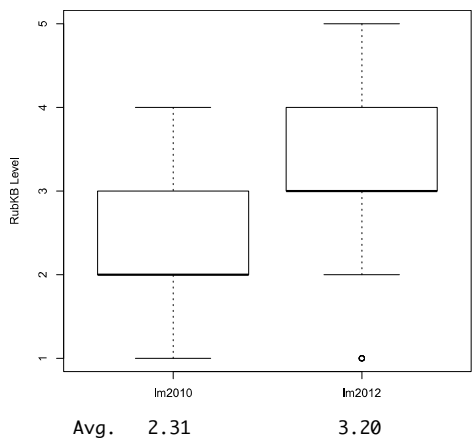

*Figure 2: The results of the question in term report "what kind of activity is ideal in group work"*

A paired t-test has been conducted for comparison. There were no significant differences between the mean scores of before and after class for the preference of general learning (t(44) = 0.66, p>.10). Whereas for the mean of after class is significantly higher than that for before class for the preference of collaborative learning (t(44) = -2.95, p<.01).

## DISCUSSION

The results indicate the effects of our trial in 2012 which we designed using the discourse analysis with KBDeX in the meta learning activities. We considered the design led to the transformation of learning beliefs from "just experiences of the participation of the collaborative learning" to "active contribution for the collaborative knowledge creation". The claim was reinforced by the results of the average score of rated students' descriptions regarding "what kind of activity is ideal in group work", indicated a higher level of understanding concerning collaborative knowledge building for the experimental group. It is difficult to conclude that students' experiences of KBDeX were the only factor of the improvement. However the only essential difference between the two groups was one group used the tool and the other did not, therefore we confidently claim the course to develop the group level of meta cognition with the tool contributed to the improvement.

The results also indicate that we succeeded to improve the students' beliefs regarding collaborative learning. Although we could not compare the results with 2010, the teacher who managed the course felt that there was great improvement. The result of the average score of the preference for general learning was 3.75/5.00 before taking the class. We did not consider this to be a low score. Whereas the score for the collaborative learning was 2.82, we recognize this is considered a low score although given the experience of Japanese university teachers with their students this score is not unusual. Experienced teachers would agree that only using group activity experiences for students cannot bring about the significant improvement of the preference regarding collaborative learning, without reflections to develop the group level of meta cognition. Hence, we claim this research demonstrated the future educational standard which is enabled by the software tool.

The limitation of this research is not involved in the evaluation of the knowledge creation outcomes. Our next step is to capture the change of the knowledge creation after doing discourse analysis by students themselves.

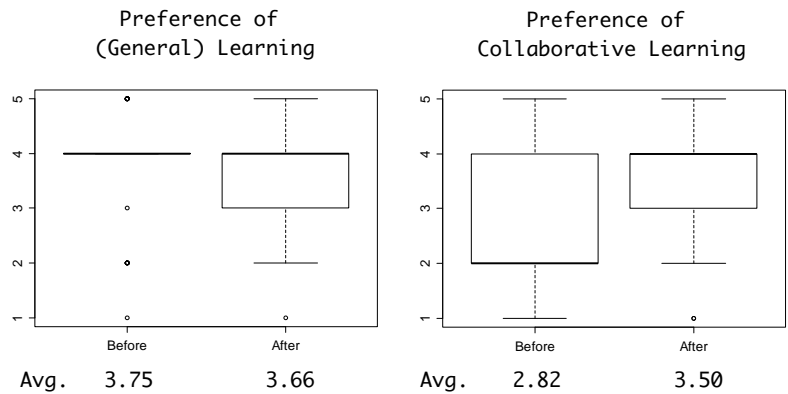

*Figure 3: The results of the questionnaire about preference about learning*

# APPENDIX

| ID | Category | Explanation, Coding Criteria | Example |
|---|---|---|---|
| SHR | Knowledge Sharing | Descriptions made by the student about the importance of knowledge sharing in collaborative learning | I think this course focused more on collaborative group learning than individual learning. This course requires argumentation by a group, integrating individual opinions. I realized that the group activity let us share various ideas. This was my first successful experience with collaborative learning in the practical situation. |
| COM | Communication Skills | Descriptions made by the student about the importance of each individual communication skill to drive conversations and to keep (not to destroy) the "good" conversation phenomena | I have found individual communication skills are important. Especially, to develop a "good" phenomena at the first meeting is critically important. The failure of development leads to forming a "bad" (inactive) group where no progression of tasks can be expected. |
| DIV | Idea Diversity | Descriptions made by the student about the findings of the importance of diversity of ideas in the collaborative learning. | I learned importance of group activities. I had thought people could learn enough individually, and I did not like group activities so much. However, the experience in this course changed my thought. Actually, I thought it was a good thing to take various ideas. Although when a discussion tended to be stuck or to focus only one side or in an individual activity, the diverse ideas in the group provided learners various aspects of the discussion topic. |
| CNT | Controversy | Descriptions made by the student about the effectiveness of collaborative learning especially for controversial topics. | I agree with the application of group learning as one of the alternatives to make an argumentation. I thought to answer the problem individually as long as possible, and I had thought that it was an advantage in my learning. However, classes in university require making an argument, whereas classes through high school required solving a quiz efficiently and accurately. I realized that group learning is effective for controversial topics, because the group discussion makes better solutions than individuals do. |

*Table A1: Coding category and criteria for report qualitative analysis for the question of "What is the most impressive thing to change your learning belief" (Part I)*

| ID | Category | Explanation, Coding Criteria | Example |
|---|---|---|---|
| ELB | Argument Elaboration | Descriptions made by the student about how collaborative learning affects the elaboration of their argumentation. | I had thought that collaborative learning was not effective due to wasting time for duplicate discussion and ideas. However, I have found some group members contributed ideas better than mine, or ideas I did not have, in my experience of group discussion in this class. We could elaborate our argument by summarizing these ideas. Although I still think that individual learning is effective, the ideas produced by that process is dangerous. |
| DEP | Deep Understanding | Description made by the student about how collaborative learning encourages deep understanding for each individual. | I had thought learning is to input and possess knowledge. But my thinking has changed to learning is (1) to individually create a hypothesis, (2) to examine it, (3) to reflect on it and improve ideas for the next activity, and (4) gain more deep understanding by repeating the process. Group learning encourages learning by improving my ideas by integrating others' ideas, and theirs' improve as well. |
| EVD | Reasoning and Evidences | Descriptions made by the student about the importance of reasoning supported by evidences. | I have learned that reasoning should be supported by evidence to convince others in collaborative learning. Others did not accept my ideas if not supported by evidences based on facts or objective data. I strongly felt that providing ideas with reasoning is important for contribution of the group. |
| CRA | Knowledge Creation | Description made by the student about how learning images have been changed into "knowledge creation" for the community. | Although my image of learning was to accumulate knowledge in my memory like learning in high school, now I think learning should be driven by intellectual interests in university. Then we could learn through making poster presentations to provide our improved knowledge. The most impressive thing I found that I can gain ideas by learning to contribute to the community. |
| PTA | Passive to Active | Description made by the student about how their thought has been changed from a passive learning style to an active learning style. (The student realized the importance of output-oriented learning) | I had thought that learning was only collecting knowledge and memorizing facts individually. But I realized that such learning is very limited in capacity for solving various types of problems. Additionally, I realized that my ideas can be improved by creating an output through the group, or I can gain ideas from other aspects of the problem. I can internalize knowledge through the process. |
| MET | Meta Learning | Description made by the student about the importance of the meta learning activities including knowledge building discourse analysis. | One of the most important experiences for me was knowledge building discourse analysis. I have never done any analysis as we conducted in this class. The sheet used in the analysis successfully guided my analysis by requiring the discussion summary, important keywords, important nodes, phase analysis, and contribution of each participant. I was so impressed that a lot of findings were found in the analysis which I could not have found in the discussion process. |
| MNG | Collaboration Management | Descriptions made by the student about collaboration management such as scheduling or task allocation. | As I have had few experiences before participating in this class, I found the importance of conducting a group project. I wondered how we avoid wasting time in the group discussion, or how we can make schedule to conduct a group project efficiently. |

*Table A2: Coding category and criteria for report qualitative analysis for the question of "What is the most impressive thing to change your learning belief" (Part II)*

| Level | Description |
|---|---|
| 1 | Students are NOT required to **work together** in pairs or groups. |
| 2 | Students DO **work together** BUT they DO NOT have **shared responsibility**. |
| 3 | Students DO have shared responsibility BUT they ARE NOT required to make substantive decisions together. |
| 4 | Students DO have shared responsibility AND they DO make substantive decisions together about the content, process, or product of their work. |

*Table A3: "Collaboration rubric" in ITL LEAP 21*

| ITL level | RubKB Level | Name | Description | Example |
|---|---|---|---|---|
| 1 | 1 | not ready for work together | Students DO NOT refer how to **work together** in groups (just refer how to start group work), | I think the members in an ideal group have their own opinions about the topic. |
| 2 | 2 | task role sharing | Group work requires DOING **work together** (only task role sharing), BUT DOES NOT require **shared responsibility.** | I think the characteristic of an ideal group is smooth communication in each group through this class. |
| 3 | 3 | knowledge sharing | Group work requires shared responsibility which can be achieved exchanging one's own thinking, BUT it IS NOT required to make substantive decisions together. | I think an ideal group creates opportunities to share opinions which create synergy. |
| 3 | 4 | solo knowledge building | Group work requires shared responsibility, not only sharing one's own thinking but also improving one's own thinking through exchange ideas with others. BUT it IS NOT required to make substantive decisions together. | I think criticisms which cause the improvement of each opinion are important. |
| 4 | 5 | collaborative knowledge building | Group work requires shared responsibility AND to make substantive decisions together about the content, process, or product of group members' work. | All the members express their own opinions actively. In the poster making phase (final product), all the members express their own opinions and create the poster collaboratively. I think this group work could improve the group members' ideas. |

*Table A4: The rubric of collaborative knowledge building attitude (RubKB)*